\documentclass[preprint,authoryear,12pt]{elsarticle2}

\usepackage{graphicx}
\usepackage{amssymb,amsmath}

\journal{Icarus}

\begin{document}

\begin{frontmatter}


  \title{Efficiency of a wide-area survey in achieving short- and
    long-term warning for small impactors}

  \author[unipi,spacedys]{Davide Farnocchia}

  \author[spacedys]{Fabrizio Bernardi} \ead{bernardi@spacedys.com}

  \author[iaps]{Giovanni B. Valsecchi}
  
  \address[unipi]{Department of Mathematics, University of Pisa, Largo
    Pontecorvo 5, 56127 Pisa, Italy}

  \address[spacedys]{SpaceDyS, Via Mario Giuntini 63, 56023 Cascina,
    Pisa, Italy}

  \address[iaps]{IAPS, INAF, Via Fosso del Cavaliere 100, Tor
    Vergata, 00133 Roma, Italy}

\begin{abstract}
  We consider a network of telescopes capable of scanning all the
  observable sky each night and targeting Near-Earth objects (NEOs) in
  the size range of the Tunguska-like asteroids, from 160 m down to 10
  m. We measure the performance of this telescope network in terms of
  the time needed to discover at least 50\% of the impactors in the
  considered population with a warning time large enough to undertake
  proper mitigation actions. The warning times are described by a
  trimodal distribution and the telescope network has a 50\%
  probability of discovering an impactor of the Tunguska class with at
  least one week of advance already in the first 10 yr of operations
  of the survey. These results suggest that the studied survey would
  be a significant addition to the current NEO discovery efforts.
%

\end{abstract}


\end{frontmatter}


\section{Introduction}
In the design of a NEO survey there is a possible trade-off between
covering less sky to a deeper magnitude or more sky to a fainter one,
as described by \citet{T11}. In the present paper we denote the first
strategy as ``Deep Survey'', and the second one ``Wide Survey''. In
the literature, the basic idea of a Deep Survey is in \citet{M92},
while the idea of a Wide Survey is described in \citet{H95}. The
choice between the two observing strategies is driven by the goals of
the survey \citep{S02}.  According to \citet{M92}, Deep Surveys such
as the present American ones are more effective in reaching the
completeness of the NEO population as they scan larger volumes of the
Near Earth space for a fixed absolute magnitude.

As a matter of fact, \citet{Ma11} claim that more than 90\% of objects
larger than 1 km (first Spaceguard goal) have been discovered so far,
predominantly by US surveys. Nevertheless, these surveys are not
optimized for detecting imminent, relatively small impactors from 10
to 160 m diameter, which may still cause very important damages and
losses on the ground. As shown by \citet[Fig. 4]{BR02}, the energy
released by such impactors ranges from 50 to $10^5$ kT and already
\citet{M92} discusses the substantial local damage that a
Tunguska-sized impactor can inflict to a populated area.

The reason why deep surveys are not suited for imminent impactors is
that their observing strategy is to cover the same area in the sky
after a few days, and to take only a minimum number of images. This
impairs the successful identification of objects that are going to
impact within a few days. For instance, \citet{V09} prove that
Pan-STARRS would not have been able to collect enough detections to
compute an orbit for $2008TC_3$, a $\sim$ 5 m asteroid which impacted
on Earth on October 2008.

To deal with imminent impacts, a more effective strategy is the Wide
Survey, as demonstrated by \citet{H95} and \citet{T11}. This kind of
survey provides a more responsive NEO impact warning system and thus
could nicely complement the current NEO discovery and cataloging
strategy of the US programs.

In the present paper we measure the performance of an assumed Wide
Survey design through a simulation of a 100 yr time span of
operations. We deal with small impactors, i.e., those with absolute
magnitude between 22 and 28, and measure the time it takes to reach a
50\% threshold for the fraction of objects discovered with warning
time sufficient to undertake proper mitigation actions.

\section{Blind time}
An impactor can arrive from almost anywhere in the sky. In particular,
if the impactor comes from the direction of the Sun, it will be most
likely not detectable in the last days before its fall. This implies
that such an object should be discovered at a previous apparition, if
at all possible. Thus, we are led to the possibility that, after the
beginning of the operations of a survey, there is no chance for the
potential impactor to be discovered before its fall; i.e., the survey
is ``blind'' for this specific impactor. Such a situation depends on
several factors, including the orbit and absolute magnitude of the
impactor, and the parameters characterizing the survey (limiting
magnitude, sky coverage, cadence, etc.).

\begin{figure}[t]
\begin{center}
\includegraphics[width=12cm]{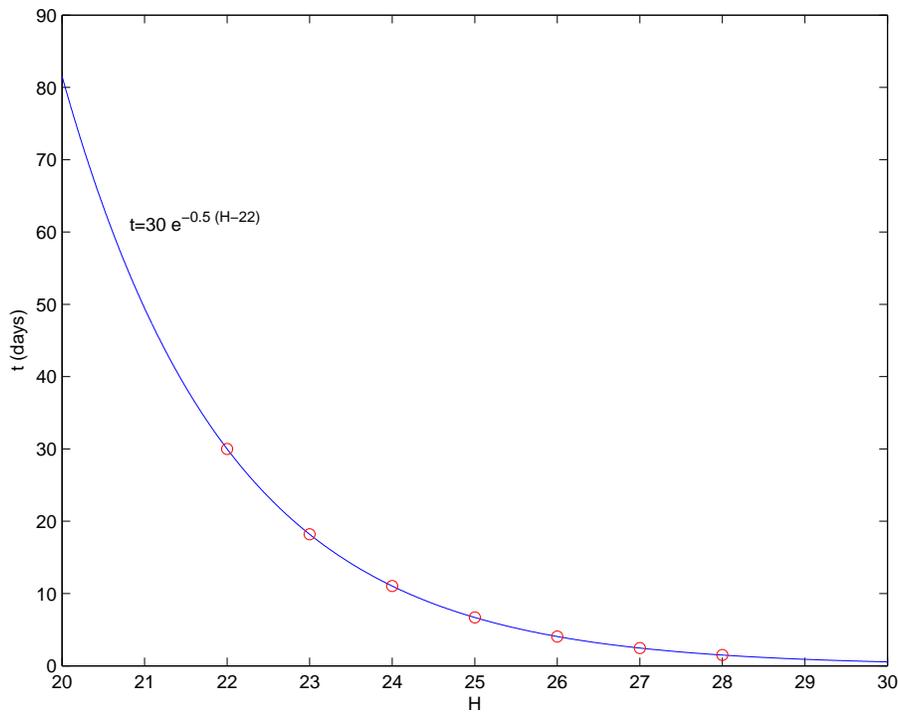}
\end{center}
\caption{Minimum lead time as a function of the absolute magnitude
  $H$.}
\label{fig:lead}
\end{figure}

Given an impactor, we define the ``lead time'' as the interval of time
between the first orbit determination and the time of
impact. According to the size of the impactor, the lead time should be
large enough to undertake the required mitigation actions, i.e., the
larger impactor the more time is necessary for the mitigation. In
absence of specific information about the albedo and the shape of an
imminent impactor, the size can be inferred from its absolute
magnitude $H$. Thus, we define the minimum required lead time as a
function of $H$, using the following constraints:
\begin{itemize}
\item a minimum lead time of 30 days for objects of $H=22$;
\item a minimum lead time of about one week for Tunguska-sized
  impactors ($H=24.5$).
\end{itemize}
We adopt the following function
\begin{equation}\label{eq:lead_time}
  t(\text{d})= c_1 e^{-c_2(H-22)}\ ,\ c_1=30\ \text{d}\ ,\ c_2=0.5\ .
\end{equation}
that fulfills the above constraints. Figure~\ref{fig:lead} shows $t$ as
a function of $H$.

It is important to point out that this simple law is tailored to the
population used in our simulations, with $H$ ranging between 22 and 28
(see Sec.~\ref{s:impactors}). For these objects, which are the target
of the Wide Survey described in this paper, the mitigation actions to
be undertaken are essentially orbit improvement and
evacuation. Dealing with bigger objects requires a different approach
and mitigation strategy, thus Eq.~(\ref{eq:lead_time}) should be
replaced with a different model, possibly involving a larger number of
parameters.

Given an NEO population, we define as ``blind time'' for a given
survey and a given absolute magnitude $\bar H$, the time between the
start of operation of the survey and the moment at which 50\% of the
impactors with magnitude $\bar H$ have a lead time larger
than the minimum threshold defined by Eq.~(\ref{eq:lead_time}). The
blind time can be used as an indicator of the performances of a given
survey. As a metric, the blind time is a variant of the time required
for a survey to discover 90\% of a defined population. The usefulness
of this definition is that, when dealing with small but numerous NEOs,
the time scale for a 90\% completion is very long and uncertain, due
to the poor modeling of the small object population.

\section{The simulation}
Hereafter we describe our assumptions on the optical network, on the
impactor population, and the orbit determination process used in the
simulation.

\subsection{Optical network}
For the optical sensors we assume the use of the innovative fly-eye
telescope design, having the following main
characteristics:
\begin{itemize}
\item an equivalent aperture of 1 m;
\item a FoV of 45 deg$^2$ ($6.66^\circ \times 6.66^\circ$);
\item high efficiency CCDs (80-90\%) with very fast read-out times
  ($\simeq 2$s) and very good cosmetics ($\simeq 99\%$);
\item a fill-factor $\simeq 1$, that is the ratio of the effectively
  detected area of the FoV and the FoV;
\item a minimum elevation of 15$^\circ$ above the horizon.
\end{itemize}
The above assumptions on the sensor hardware require a significant
effort in both technological development and resources. The concept
design of the assumed telescope is described in \citet{C11}. However,
we are aware that these assumptions need to be validated by further
studies when the telescope is actually available, at least in a
prototype phase. A discussion on this is beyond the scope of this
paper.

For the telescope network we assume:
\begin{itemize}
\item One equivalent dedicated survey telescope in the northern and one
  in the southern hemisphere.  
\item The northern telescope covers the northern hemisphere of the
  celestial sphere, while the southern telescope covers the southern
  hemisphere.
\item One dedicated follow-up telescope in the northern and one in the
  southern hemisphere, typically 30$^\circ$ West of the survey
  telescopes.
\item The images are processed locally in real time, included the
  astrometric reduction, and the data are made available to the
  scientific community in less than two hours. Therefore, the
  dedicated follow-up telescopes can be triggered to follow the newly
  discovered objects.
\end{itemize}
With these assumptions each equivalent telescope can take about 766
images for an average 10 hour night. This corresponds to a total of
about $34450$ deg$^2$ which is equivalent to $17225$ deg$^2$ of the
celestial sphere taken twice per night.

For the observing strategy we assume:
\begin{itemize}
\item Observations that cover $36400$ deg$^2$ ($\simeq 88\%$ of the
  celestial sphere), corresponding to all the visible sky except the
  regions with solar elongation less than $40^\circ$.
\item The regions of the sky within $30^\circ$ of the Moon or within
  $15^\circ$ of the galactic plane are not covered by the telescopes
  due to the increase of the sky background. Therefore, the effective
  visible sky ranges between 22987 deg$^2\simeq 56\%$ of the celestial
  sphere (when the forbidden regions around the Sun, the Moon and the
  galactic plane do not overlap) and 34348 deg$^2\simeq
  83\%$ (when the intersection between the forbidden regions is
  maximized). On average, each telescope covers between $11500$ and
  $17200$ deg$^2$.
\item A limiting magnitude $V_{lim}=21.5$, corresponding to $\simeq 45$
  s of exposure time, for the survey mode, and $V_{lim}=23$ for the
  follow-up mode.
\item Coverage of the visible sky at least two times per night. 
\end{itemize}

In a real system the number of telescopes has to be increased and
could be between 5 and 6. Indeed, to deal with the cloud coverage and
meteorological correlations we need a minimum of two survey telescopes
per hemisphere widely spread in longitude. Furthermore, to increase
the detection efficiency, a higher number of detections may be
necessary and also this can be achieved with a higher number of survey
telescopes.

\subsection{Impactor population}
\label{s:impactors}
\begin{figure}[h]
\centerline{\includegraphics[height=5.5cm]{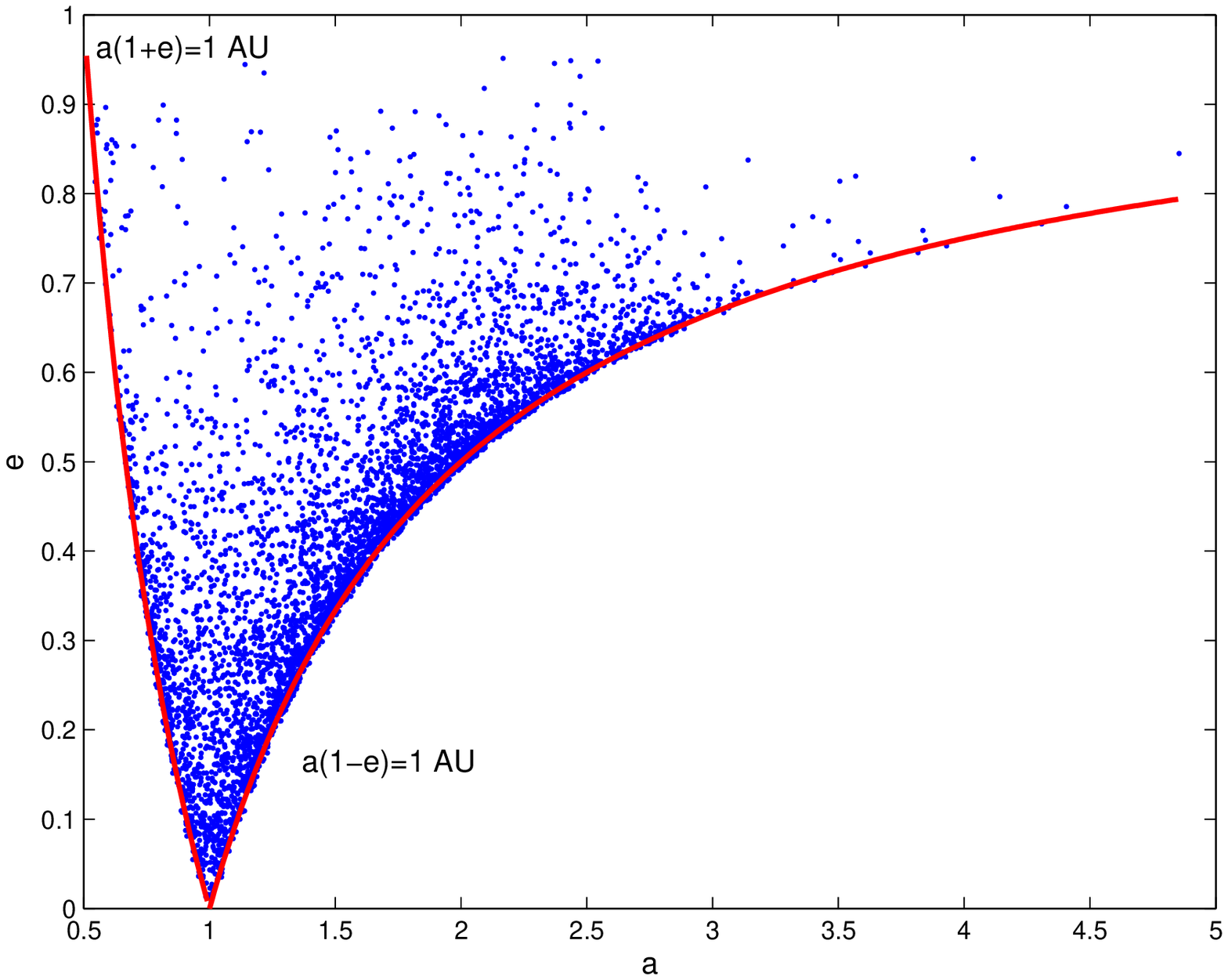}
\includegraphics[height=5.5cm]{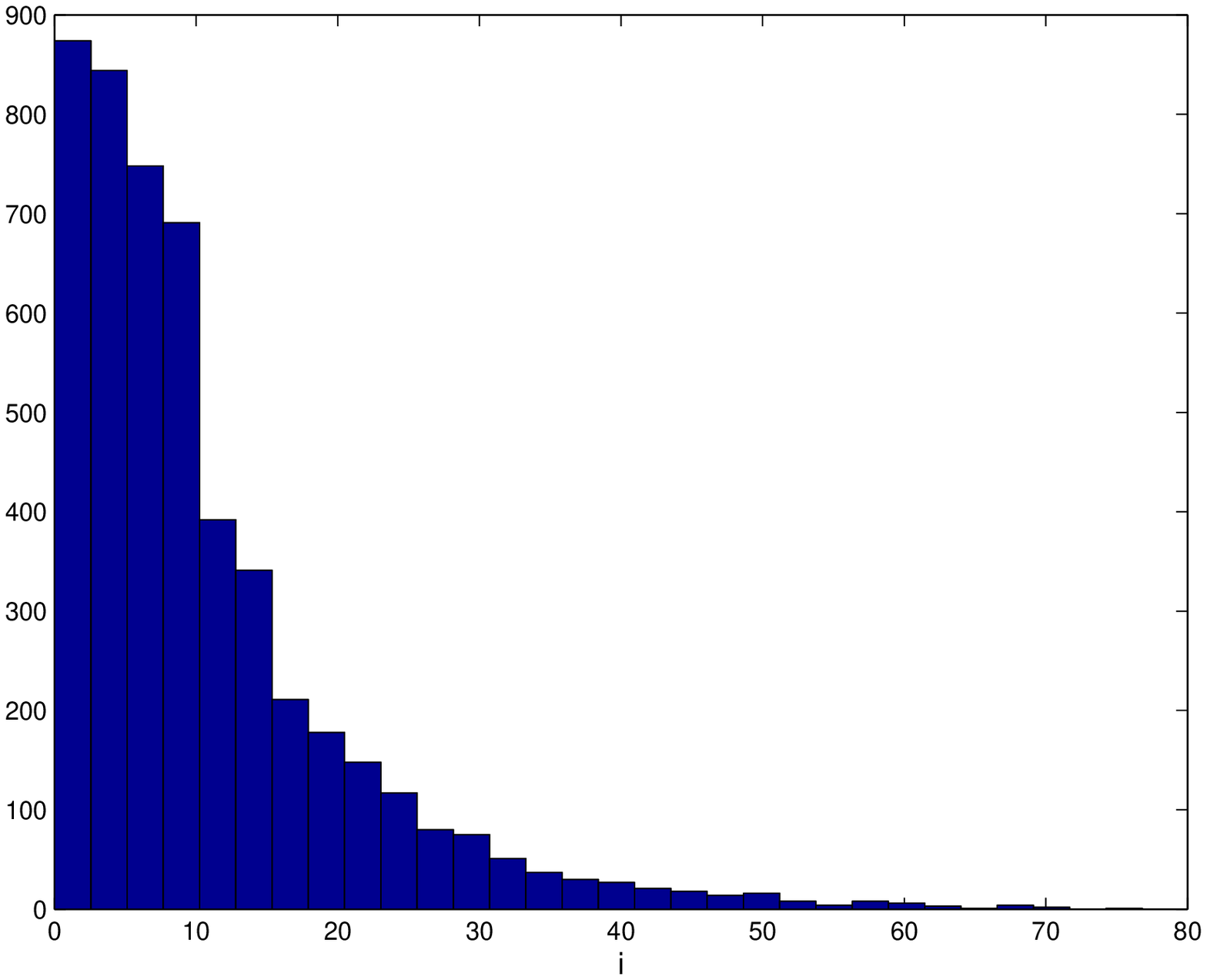}}
\caption{Left: scatter plot in the $(a,e)$ plane of the impactor
  population by \protect{\citet{C04}}. The two solid lines enclose the
  region of Earth crossing orbits. Right: the distribution of
  inclinations of the same population.}
\label{fig:aei}
\end{figure}

In our simulation we use the population of 4950 synthetic impactors
described by \citet{C04} which have impacts in a time frame of 100 yr
starting from July 2009. This impactor population is selected within
the population model by \citet{B02}. Figure~\ref{fig:aei} shows the
distribution of semimajor axis, eccentricity and inclination. The
majority (68\%) of the objects has a perihelion between 0.8 and 1 AU.

We assign a fixed value of the absolute magnitude $H$ to all the
asteroids and we repeat the simulation for integer values of $H$
ranging between 22 and 28, roughly corresponding to diameters between
160 and 10 m. We choose this simulation strategy to measure the
performance of the proposed network as function of the size range of
the asteroids.

\begin{figure}[t]
\begin{center}
\includegraphics[width=12cm]{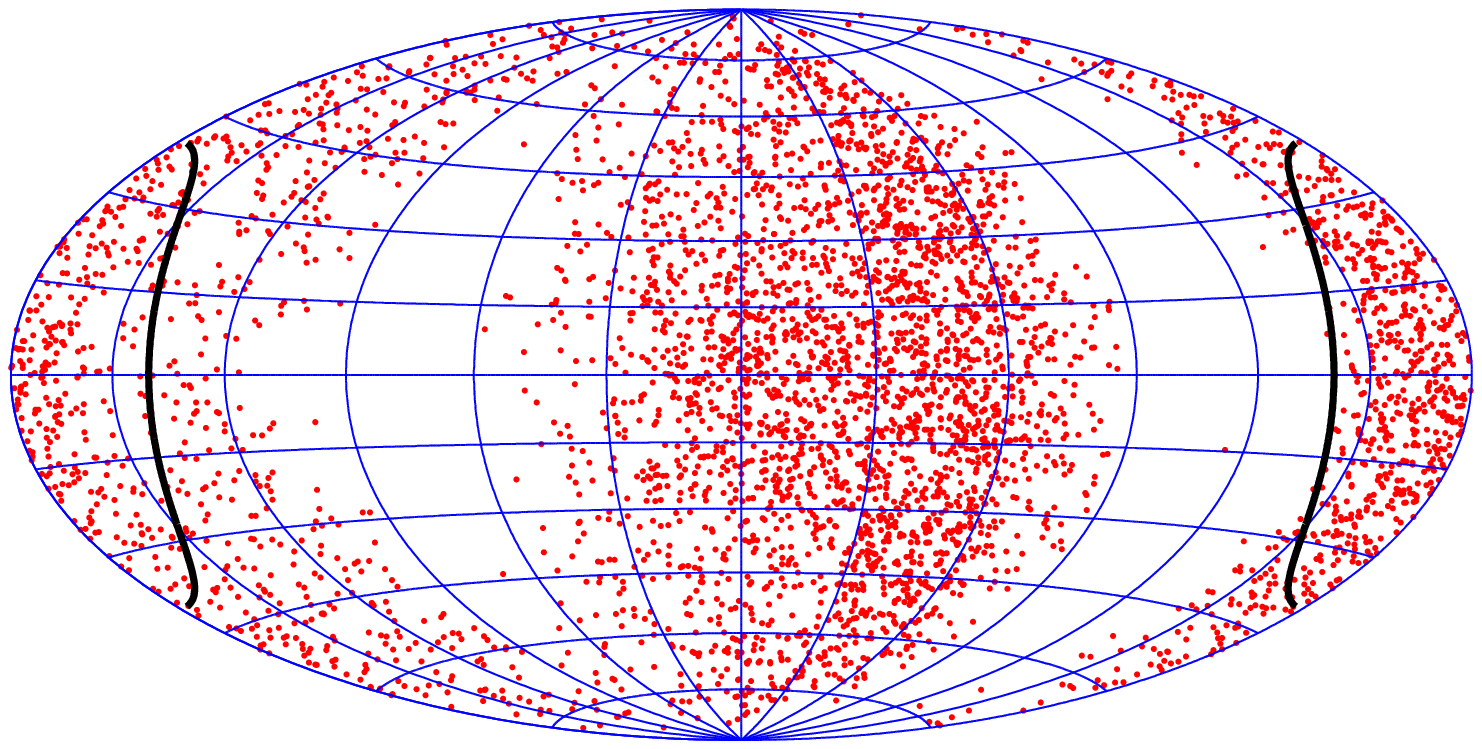}
\end{center}
\caption{The radiant distribution of the simulated impactors of
  \citet{C04}. The radiants are shown in an equal area projection of
  the sky centered on the opposition; the angular coordinates are
  ecliptic longitude minus the longitude of the Sun, and ecliptic
  latitude. The bold lines refer to 40$^\circ$ of solar elongation.}
\label{sky}
\end{figure}
To obtain the sky distribution of the impactors shortly before the
event, it is useful to plot their radiants (see \ref{s:rad}).
Figure~\ref{sky} shows the distribution of the radiants for 4\,465
Earth impactors in such a representation\footnote{The analytical
  procedure to compute the radiants assumes a circular orbit for the
  Earth. 485 among the objects of \protect{\citet{C04}} have either
  $a(1-e)>1$ or $a(1+e)>1$, so that they are excluded from the
  analytical computation.}. The sky distribution of impactor radiants
is far from uniform and is a consequence of the $a$-$e$-$i$
distribution of the impactor population. The fraction of radiants with
a solar elongation larger than $40^\circ$ ---the minimum elongation
from the Sun at which the assumed survey can observe--- is $80.1\%$ of
the whole sample. It is worth noticing that, for the impactors with
radiant within 40$^\circ$ of the Sun, a detection at an apparition
before the one corresponding to the impact is the only chance to have
a long lead time.

\subsection{Methodology}
We split the impactor population in 10 bins according to the impact
epoch with respect to the beginning of the simulation. For example,
the first bin contains objects impacting within 10 yr, the second one
objects impacting between 10 and 20 yr, and so on. Such a binning
allows us to measure the performance as a function of the time from
the start of the survey.

For each object we generate a list of observations according to the
assumed configuration, the performance of the optical network, and the
visibility constraints. The simulation provides one tracklet (see
later) per night for survey telescopes and up to two tracklets per
night for follow-up telescopes. The follow-up observations are
triggered once the object has been detected by the survey
telescope(s), with a minimum delay time of two hours. For each
simulated observation Gaussian noise with a standard deviation of 0.3
arcsec is added to the astrometric position. Similarly a Gaussian
noise is added to the magnitude estimate for the detection, but in
this case the noise is split in a correlated component (for the light
curve effect) of 0.2 magnitudes and a random component, again of 0.2
magnitudes.

The tracklet is the atomic information for a moving asteroid,
consisting of a small number (2-5) of detections in different images
of the same field, taken at moderately short intervals of time (15 min
to 2 hours).  A tracklet normally provides an amount of information
which can be described by 4 scalar quantities (two angles and two
angular rates), therefore such detections do not imply discovery
\citep{M07}. We consider as discovered a moving object belonging to
the solar system only when enough information has been accumulated to
establish its dynamical properties, that is by means of a heliocentric
orbit, for which at least 6 scalar quantities are required.

The orbit determination process starts by selecting n-tuples of
tracklets which could belong to the same object. Then, for each of them
a preliminary orbit compatible with all the tracklets is computed,
using the methods described in \citet{M04} and \citet[Chapters 7 and
8]{M10}.  Thereafter, the preliminary orbit is used as first guess in a
differential correction procedure, which usually converges to a least
squares fit orbit. If the orbital fit satisfies suitable quality
control conditions, this can be considered a real object.

To perform the simulation we set up a data center architecture, by
ingesting observational data day by day. Each time new observations
become available we update the previously known orbits and compute the
new ones, corresponding to newly discovered objects.

It is important to point out that a more realistic simulation should
take into account the presence of Main Belt background asteroids,
which increases the computational load and the rate of occurrence of
false identifications. However, observations of known Main Belt
asteroids should be filtered before looking for new objects, as
discussed in \citet{M12}.

For completeness, our procedure could include the risk assessment for
each simulated impactor including the explicit computation of an
impact probability. Such a procedure would require a very large amount
of CPU time and was performed in a previous study \citep{F11} using a
subsample of the present population over the first 20 yr of survey
operations. That paper shows that, in more than $99.5\%$ of the cases,
the availability of an orbit involving at least 4 tracklets for a
newly discovered potential impactor is accompanied by the successful
computation of an impact probability by using the CLOMON2 software
robot \citep{M05}. Such a percentage is so high as to make not cost
effective the computation of the entire impact monitoring chain in the
present case. Thus, we stipulate that an impactor is considered as
discovered when an orbit from at least 4 tracklets is computed.

\section{Results and Discussion}
\begin{figure}[h]
\centerline{\includegraphics[width=7cm]{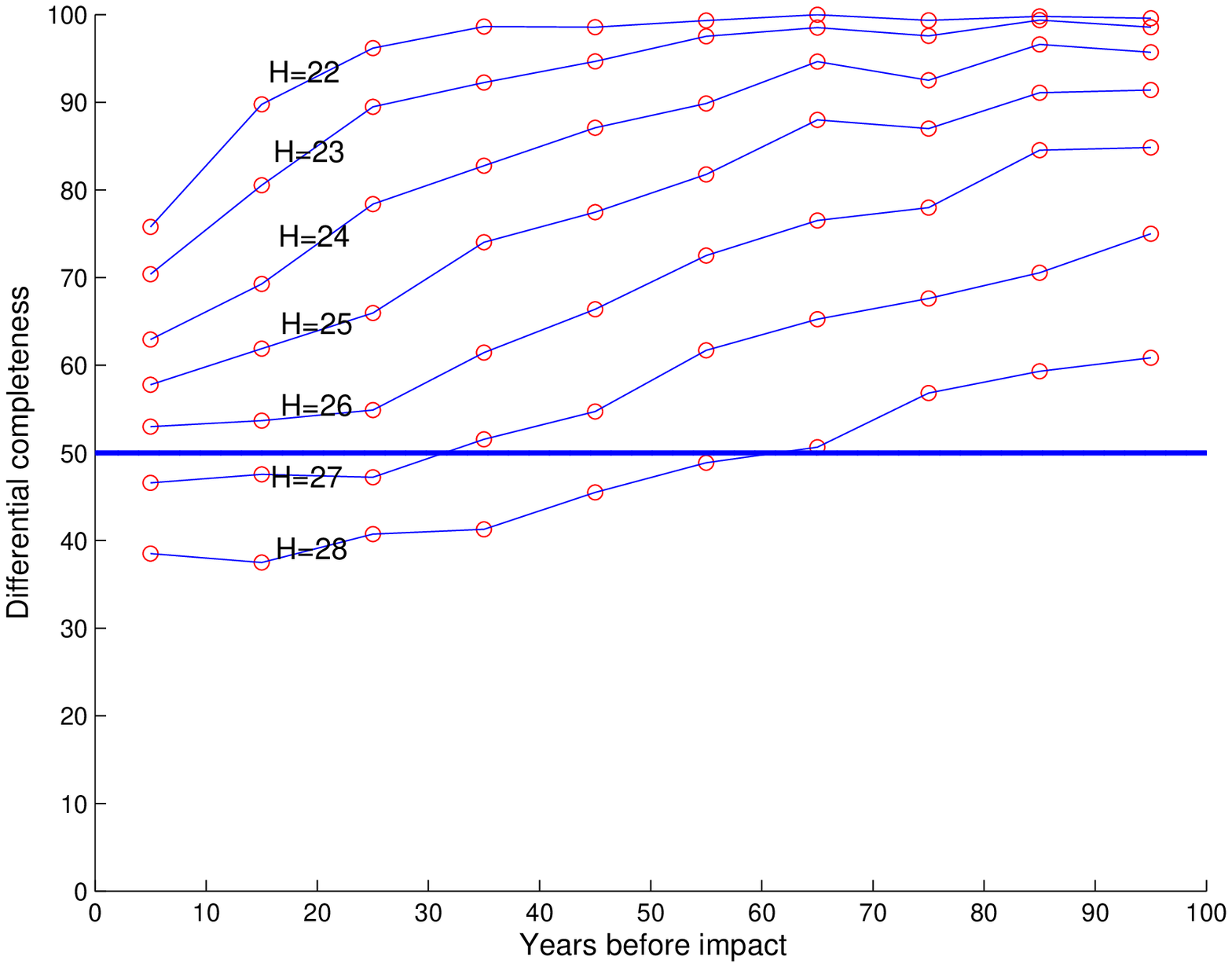}
  \includegraphics[width=7cm]{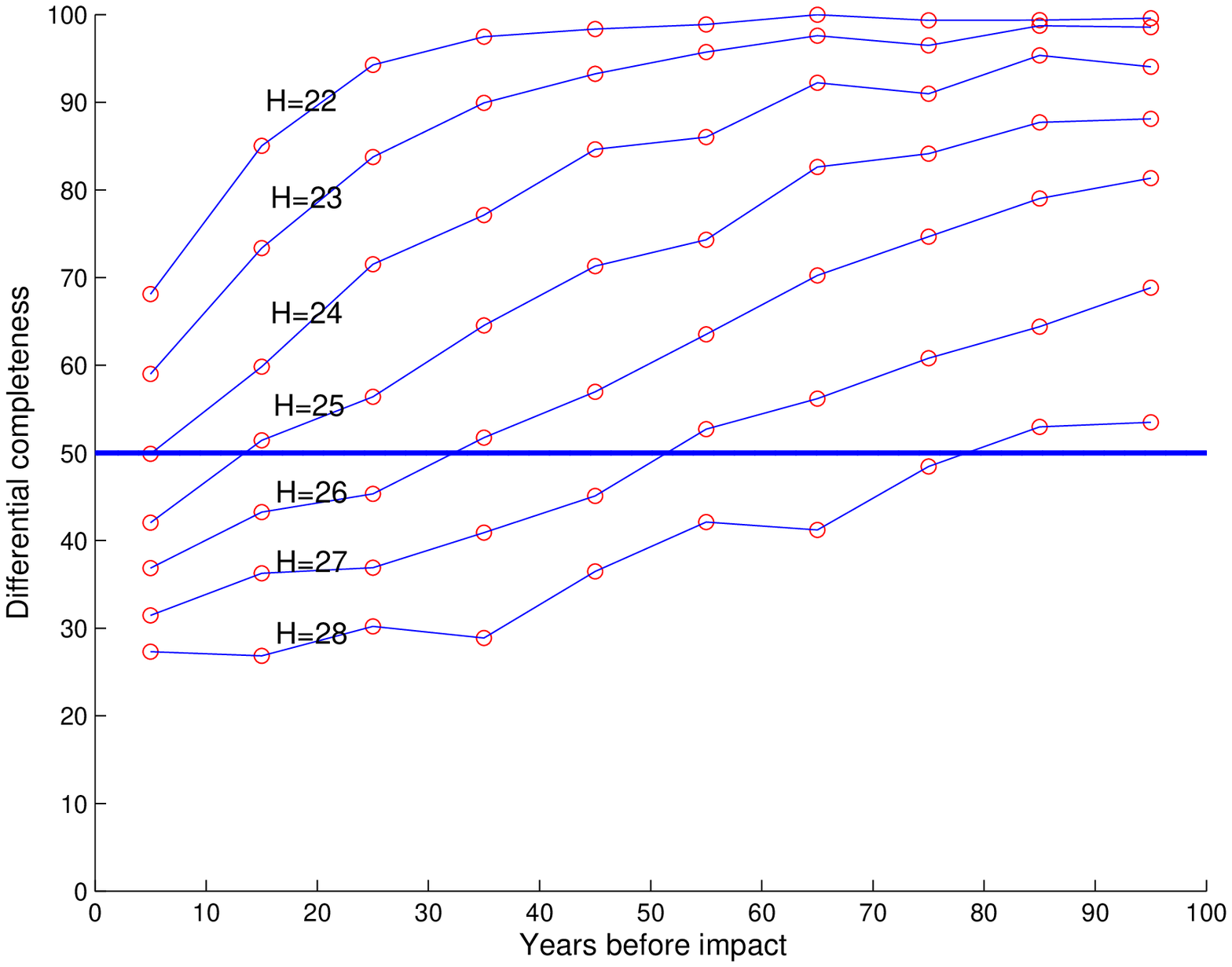}}
\caption{Left: differential discovery completeness as function of the
  impact date and absolute magnitude. Right: same as left panel,
  considering successful a discovery with lead time greater than the
  minimum threshold defined by Eq.~(\protect{\ref{eq:lead_time}}).}
\label{efficiency_diff}
\end{figure}

The main outcome of the simulation is shown in
Fig.~\ref{efficiency_diff}. 
The left panel shows the differential discovery completeness as a
function of time from the survey beginning and absolute magnitude.
For each 10 yr bin, the differential discovery completeness is
defined as the ratio between the number of impactors discovered and
those impacting in that time frame. For example, the differential
discovery completeness for $H=24$ and the bin from 20 to 30 yr is given
by the fraction of objects discovered before the impact among those
impacting between 20 and 30 yr with respect to the beginning of the
survey operations.

For impactors with $H=23$ a 70\% discovery completeness is achieved in
the first decade, while the 90\% threshold is reached after two more
decades.  For impactors with $H=25$ a 60\% discovery completeness is
achieved after two decades, while the 90\% threshold is almost reached
at the end of the simulation. For $H=28$ the discovery completeness
starts slightly below 40\%, and increases slowly during the next
decades as expected, due to the small size of the objects. Notice that
the completeness is greater than $50\%$ already from the start for
$H=26$.

The right panel shows the differential completeness for a lead time
given by Eq.~(\protect{\ref{eq:lead_time}}). This means that a
discovery is considered successful only if it takes place sufficiently
early, allowing the mitigation actions appropriate for the size of the
impactor.  For impactors with $H=23$ a $\simeq60\%$ completeness is
achieved in the first decade, while the 90\% threshold is reached
after 30 yr. For $H=25$ a $\simeq 50\%$ discovery completeness is
achieved in the second decade, while the 90\% threshold is not yet
reached at the end of the simulation. For smaller asteroids the
discovery completeness is larger than 25\% from the beginning of the
operations and overcomes the 50\% threshold after more than 70 yr.

We can sum up the results in terms of blind time, that is the
intersection between the 50\% completeness and the $H=const$ curves in
Fig.~\ref{efficiency_diff}. The survey simulated in this paper would
have a blind time of about 20 yr for imminent impactors with
$H=25$. Note that a Tunguska-sized ($H\simeq24.5$) object impacting
within 10 yr from the start of the survey would have a $>60\%$
probability of being discovered and would have a lead time larger than
1 week with a probability $\simeq45\%$. For smaller impactors the
blind time increases up to $\sim$60 yr for $H=28$.

\begin{figure}[ht]
\centerline{\includegraphics[width=7cm]{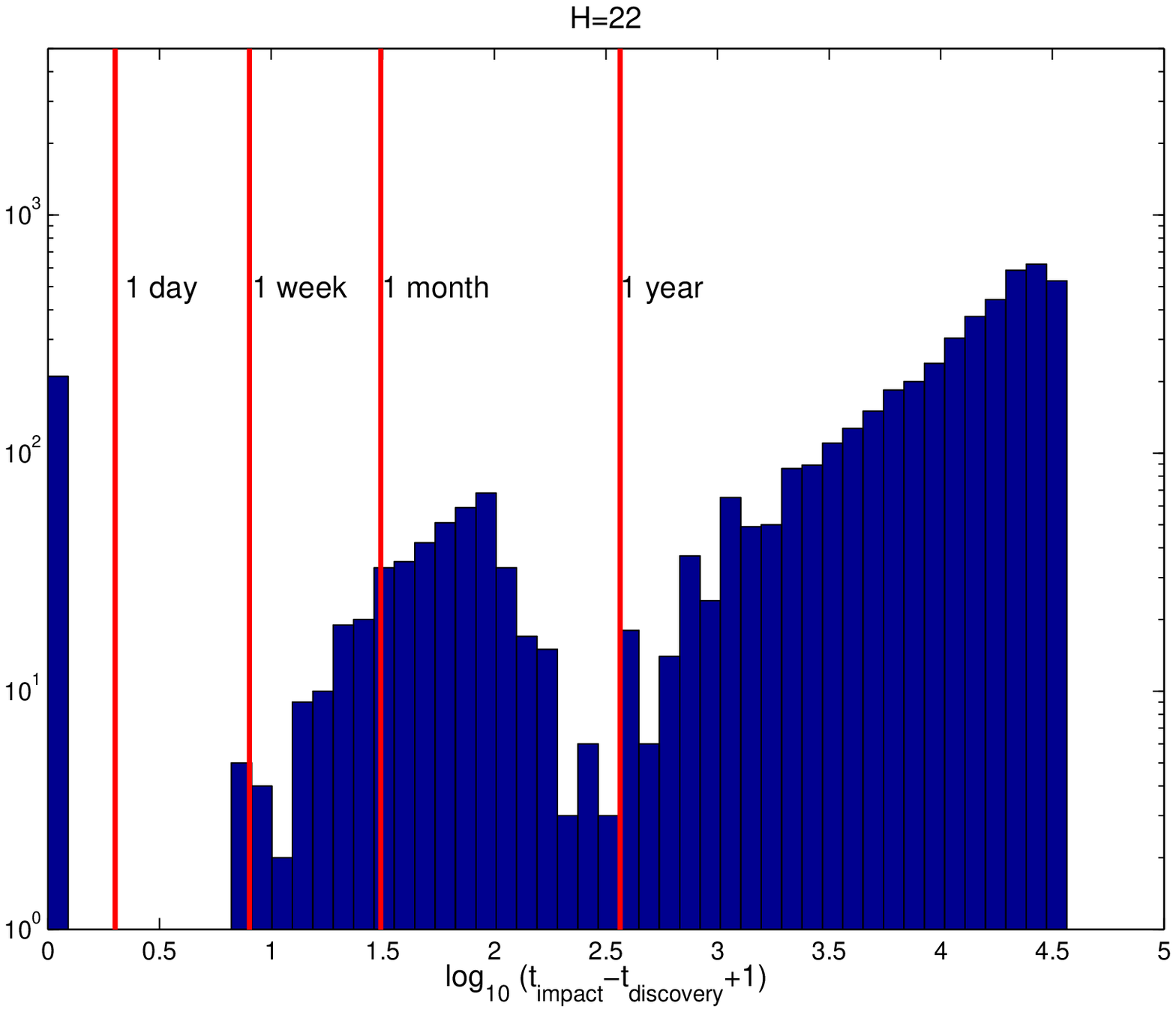}
\includegraphics[width=7cm]{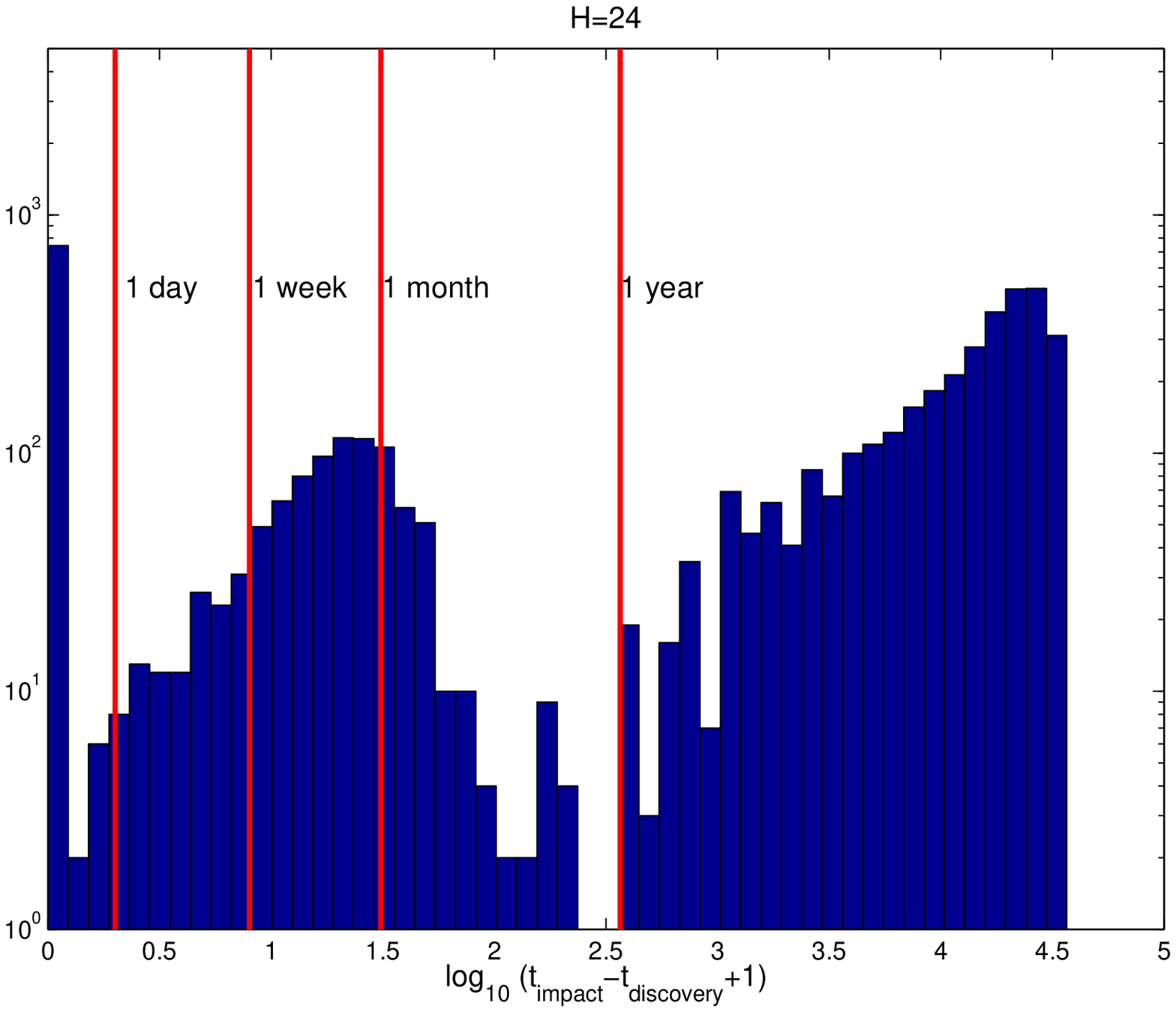}}
\centerline{\includegraphics[width=7cm]{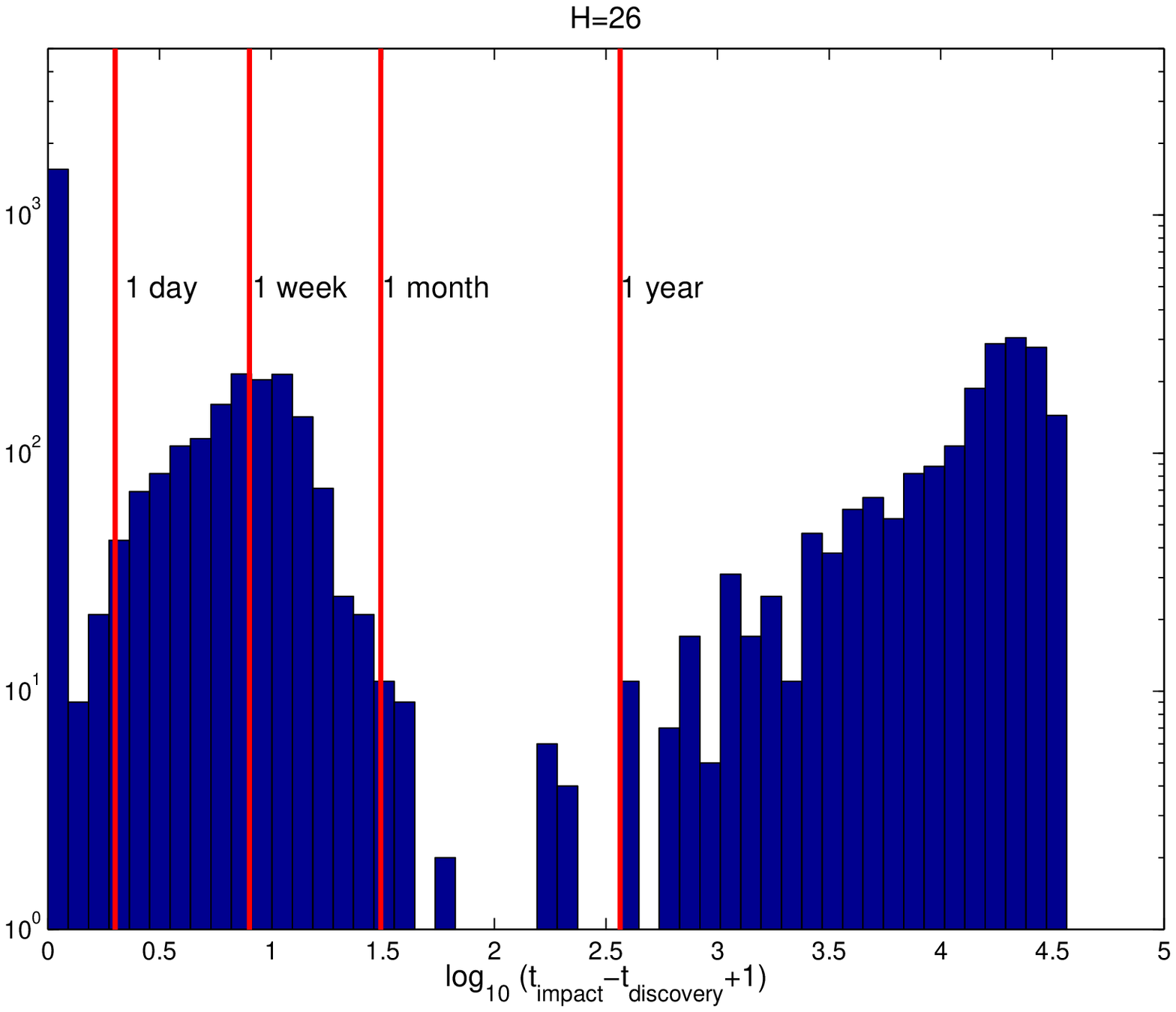}
\includegraphics[width=7cm]{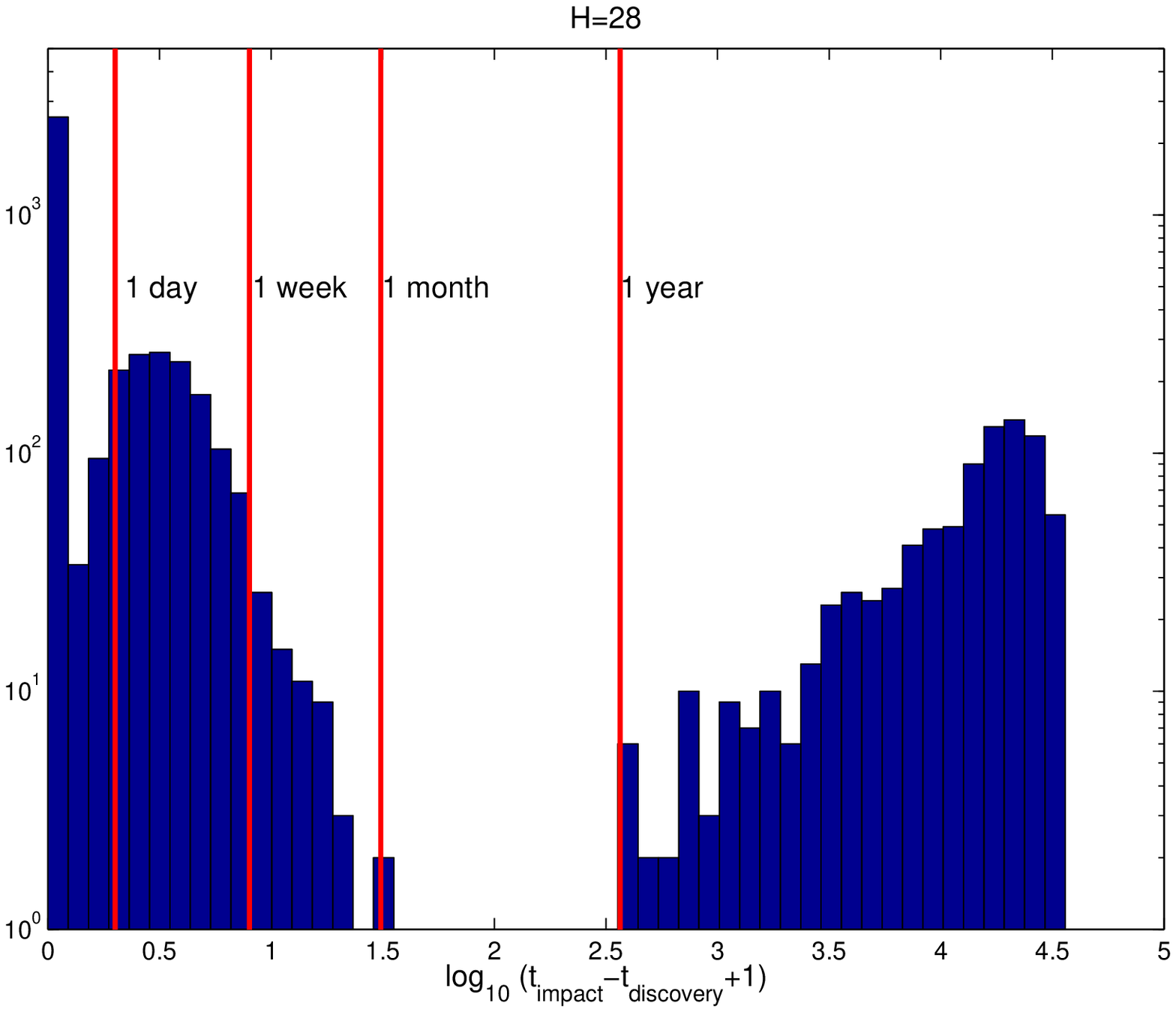}}
\caption{Histogram of the lead times for different values of $H=$22
  (top left), 24 (top right), 26 (bottom left), 28 (bottom right). The
  vertical lines denote, from left to right, 1 day, 1 week, 1 month,
  and 1 yr. Note that the scales are different.}
\label{wt_det}
\end{figure}

Figure~\ref{wt_det} shows the distribution of the lead time for
different values $H$. As expected, the lead time strongly depends on
the value of the absolute magnitude. A clear trimodality is visible in
all the panels: either the object impacts without being discovered
(left bar), or is discovered during its last apparition (central
peak), or at a previous apparition (right peak). Table~\ref{peaks}
details the fractions of objects in each peak for a fixed value of
$H$. Most of the impactors with $H=22$ are discovered during a
previous apparition with respect to the impact. As $H$ increases there
are more and more cases of objects either discovered during the last
apparition or not discovered at all.

\begin{figure}[h]
\centerline{\includegraphics[width=7cm]{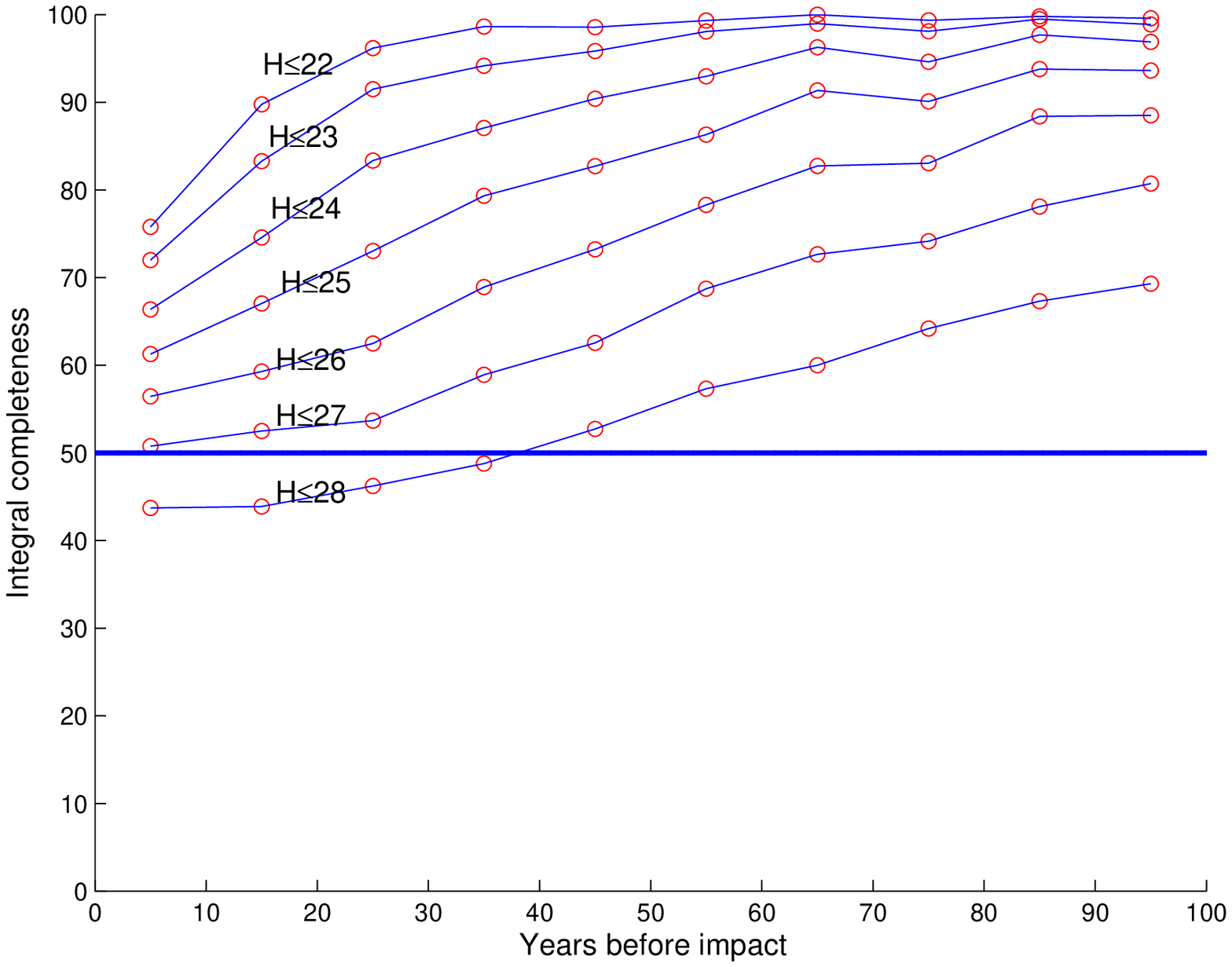}
  \includegraphics[width=7cm]{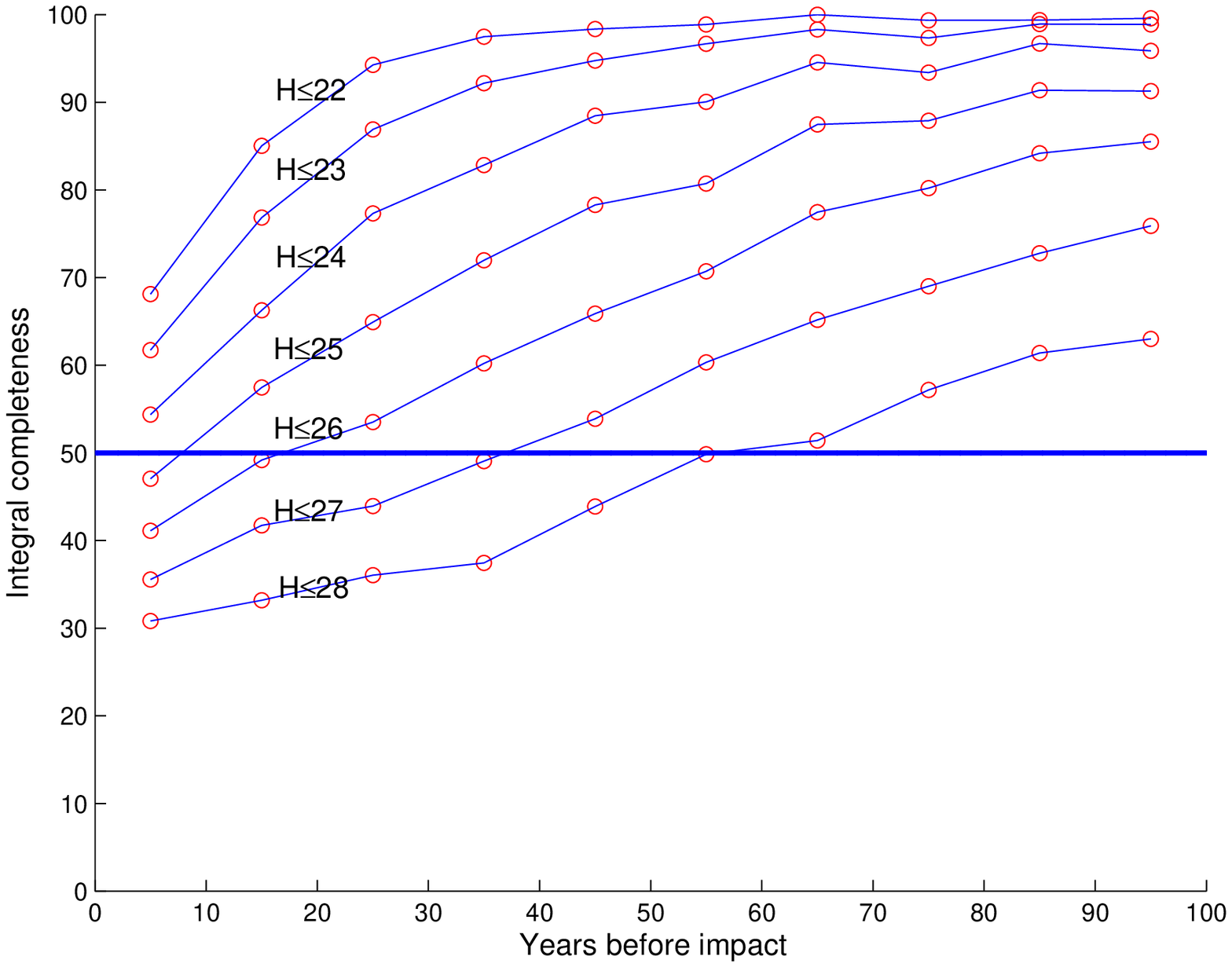}}
\caption{Left: integral discovery completeness as function of the
  impact date and absolute magnitude. Right: same as left panel,
  considering successful a discovery with lead time greater than the
  minimum threshold defined by Eq.~(\protect{\ref{eq:lead_time}}).}
\label{efficiency_int}
\end{figure}
To conclude this discussion we report the integral completeness
achieved by our simulated survey.  The integral completeness is
computed by the weighted sum
\[
\text{Comp}(H\leq \bar H)=\sum_{22\leq H_i\leq\bar H}w_{H_i}
\text{Comp}(H_i)\ \big{/} \sum_{22\leq H_i\leq\bar H}w_{H_i}
\]
where $\text{Comp}(H)$ is the completeness for a fixed absolute
magnitude $H$. To get a more realistic result we keep into account a
power law distribution for the number of asteroids at a given absolute
magnitude. Consistently with \citet{B02} and \citet{S04}, the computed
efficiencies for a fixed absolute magnitude value $H$ are given the
weights:
\[
w_H=10^{0.37(H-28)}\,.
\]
The results are summarized in Fig.~\ref{efficiency_int}.

\begin{table}[t]
\begin{center}
\begin{tabular}{r|r|r|r}
$H$ & undiscovered & last apparition & previous apparition\\
\hline
22 & 4.2\% & 8.9\% &  86.9\%\\
23 & 8.1\% & 13.6\% & 78.4\%\\
24 & 15.0\% & 18.5\% & 66.5\%\\
25 & 22.3\% & 25.8\% & 51.9\%\\
26 & 31.4\% & 31.0\% & 37.6\%\\
27 & 41.3\% & 33.0\% & 25.7\%\\
28 & 52.1\% & 31.1\% & 16.8\%
\end{tabular}
\end{center}
\caption{Percentages of impactors not discovered (2nd column), 
  discovered during the last apparition (3rd column) or discovered 
  at a previous apparition (4th column), as a function of $H$ (1st column).}
\label{peaks}
\end{table}

\section{Conclusions}
We simulated the operations, over a time span of 100 yr, of a Wide
Survey capable of covering all the sky at solar elongation larger than
$40^\circ$, down to apparent magnitude 21.5, with a nightly
cadence. The survey includes the operation of follow-up with a
limiting magnitude 23.0.

The goal of the simulation was to compute the ``blind time'' of the
survey, i.e., the time between the start of survey operations and the
moment at which 50\% of the impactors at a given magnitude are
discovered and their orbits determined, with an advance large enough
to allow undertaking the appropriate mitigation actions. In fact, our
modeling allowed us to compute a realistic distribution of the lead
time, i.e., the interval of time between the first orbit determination
and the time of impact. The distribution shows a trimodality
corresponding to 1) undiscovered objects, 2) objects discovered at the
impact apparition, and 3) objects discovered during a previous
apparition.

The survey discussed in this paper can efficiently deal with
Tunguska-sized impactors for which the blind time is about 10 yr. This
means that, already in the first ten years of survey operations, there
would be 50\% probability of discovering such an impactor at least one
week before impact. The pre-impact discovery of Tunguska-sized
impactors was not an original goal of past NEO surveys but the case of
2008TC3 and the present paper show that it is becoming within reach.

In a future paper, we plan to compare the performances of the
discussed Wide Survey and a state of the art Deep Survey thus allowing
us to quantitatively measure the contribution of the Wide Survey to
the current NEO search programs.

\section*{Acknowledgements}
This study was partly supported by ESA Contract n. 22929/09/ML/GLC and
by the PRIN INAF ``Near Earth Objects''. The authors wish to thank
A. Milani for useful discussions during the development of this work,
S.~R. Chesley for providing us with the impactor population.  We thank
S.~R. Chesley and an anonymous referee for their constructive
comments.


\appendix

\section{Impactor radiants}
\label{s:rad}
Following \citet{V99} it is possible to establish that in the framework
of \"Opik's theory of close encounters \citep{O76} the angles $\theta$
and $\phi$ define the direction opposite to that from which an Earth
impactor seems to arrive (the so-called radiants in meteor astronomy).
These angles can be computed from the orbital elements $a$, $e$, $i$
of the impactor ($a$ must be given in units of the orbital semimajor
axis of the planet, for Earth impactors in AU) as follows:
\begin{eqnarray*}
  \cos\theta & = & \frac{\sqrt{a(1-e^2)}\cos{i}-1}{\sqrt{3-1/a-
      2\sqrt{a(1-e^2)}\cos{i}}} \\
  \sin\theta & = & \frac{\sqrt{2-1/a-a(1-e^2)\cos^2{i}}}
  {\sqrt{3-1/a-2\sqrt{a(1-e^2)}\cos{i}}} \\
  \sin\phi & = & \pm\frac{\sqrt{2-1/a-a(1-e^2)}}{\sqrt{2-1/a-
      a(1-e^2)\cos^2{i}}} \\
  \cos\phi & = & \pm\frac{\sqrt{a(1-e^2)}\sin{i}}{\sqrt{2-1/a-
      a(1-e^2)\cos^2{i}}},
\end{eqnarray*}
where in the expression for $\sin\phi$ the upper sign applies to
collisions in the post-perihelion branch of the orbit, and in that for
$\cos\phi$ to collisions at the ascending node.

\begin{table}[t]
\begin{center}
\begin{tabular}{cc|cc}
\hline
$\sin\phi>0$ & $\cos\phi>0$ & $\omega+f=0$   & $0<f<\pi$    \\
$\sin\phi<0$ & $\cos\phi>0$ & $\omega+f=0$   & $\pi<f<2\pi$ \\
\hline
$\sin\phi>0$ & $\cos\phi<0$ & $\omega+f=\pi$ & $0<f<\pi$    \\
$\sin\phi<0$ & $\cos\phi<0$ & $\omega+f=\pi$ & $\pi<f<2\pi$ \\
\hline
\end{tabular}
\end{center}
\caption{The relationship between the quadrant of $\phi$ and the orbital 
  elements $\omega$ and $f$ of the 
  impactor at collision.}
\label{quadrant}
\end{table}

Thus, to each triple $a$, $e$, $i$ correspond four encounter
geometries, all characterized by the same value of $\theta$, that
differ for the quadrant of $\phi$; this, in turn, can be computed from
the orbital elements of the impactor, as shown in
Table~\ref{quadrant}.

Meteor radiants are often plotted using an equal area projection
centered on the direction of the Earth motion, with the ecliptic as
reference plane; in this case, starting from $\theta$ and $\phi$
computed from the meteoroid orbits, the radiant coordinates are simply
given by $\lambda+\pi$ and $-\beta$ ($\beta=0$ defines the ecliptic
plane, and $\lambda=\beta=0$ is the direction of the Earth motion),
with $\lambda$ and $\beta$ given by
\[
\sin\beta = \sin\theta\cos\phi \ \ ,\ \ \sin\lambda =
\frac{\sin\theta\sin\phi}{\cos\beta} \ \ , \ \ \cos\lambda =
\frac{\cos\theta}{\cos\beta}.
\]
For an asteroidal Earth impactor, it is useful to use the same
representation, but in this case centered on the opposition point,
something that is obtained by rotating $\lambda$ by $\pi/2$ in the
appropriate direction.

\subsection{An analytical expression}
The non-uniform sky distribution shown in Fig.~\ref{sky} can be
exploited in prioritizing the sky coverage for a survey aimed at
detecting very close Earth approachers and impactors.

In this respect, an important quantity to take into account is the
angular distance of the radiant from the Sun, that we will denote with
$\sigma$, since with ground based optical telescopes it is practically
impossible to observe at values of this quantity smaller than some
practical limit.  Thus, in order to evaluate the efficiency of an
impactor-aimed sky survey, it is important to establish what fraction
of impactor radiants lies below this angular distance from the Sun; to
this purpose, it is necessary to discuss the geometric setup of
\"Opik's theory \citet{V99}.

We use a reference frame centered on the Earth, with the $z$-axis
perpendicular to the plane of the ecliptic, the $y$-axis in the
direction of the Earth velocity and the $x$-axis pointing away from
the Sun, which is located at $x=-1$, $y=0$, $z=0$; in this frame, the
unperturbed geocentric encounter velocity $\vec{U}$ of the NEO has
components
\[
(U_x,U_y,U_z) = (U\sin\theta\sin\phi, U\cos\theta,U\sin\theta\cos\phi)
\]
where $U$ is the magnitude of the velocity vector.  With these
definitions, the cosine of the angle between $\vec{U}$ and the
$x$-axis is simply given by $\sin\theta\sin\phi$; this angle, in turn,
is equal to $\pi-\sigma$ since, as noted before, $\theta$ and $\phi$
define the direction opposite to the radiant.

Thus, if $\sigma_{min}$ is the minimum angular distance from the Sun
that the survey can reach, the radiants that are not observable by it
are characterized by
\begin{displaymath}
\cos\sigma_{min}<\sin\theta\sin\phi.
\end{displaymath}
Since for each triple $a$, $e$, $i$ we have four possible associated
radiants, two of which with $\sin\phi>0$ and the other two with
$\sin\phi<0$, the consequence of the above inequality is that for
$\sigma_{min}<\pi/2$ we have the following two cases:
\begin{itemize}
\item for $|\cos\sigma_{min}|\ge|\sin\theta\sin\phi|$, all four
  radiants associated to a given triple $a$, $e$, $i$ are observable;
\item for $|\cos\sigma_{min}|<|\sin\theta\sin\phi|$, the two radiants
  for which $\sin\phi<0$ are observable, while the other two are not.
\end{itemize}
Given a population of impactors, it is then possible to compute the
fraction $F$ of radiants that have $\sigma>\sigma_{min}$; if $h$ is
the fraction of the population characterized by
$|\cos\sigma_{min}|\ge|\sin\theta\sin\phi|$, then
\[
F = h+\frac{1-h}{2} =
\frac{1+h}{2}.
\]

\bibliographystyle{elsarticle-harv}

\end{document}